\documentclass[12pt,onecolumn]{emulateapj}

\newcommand{\be}{\begin{equation}}
\newcommand{\ee}{\end{equation}}
\newcommand{\ba}{\begin{eqnarray}}
\newcommand{\ea}{\end{eqnarray}}

\begin{document}

\title{Studying Turbulence from Doppler-broadened Absorption Lines: Statistics of Logarithms of Intensity}

\author{A. Lazarian}
\affil{Department of Astronomy, University of Wisconsin,
Madison, US}
\author{D. Pogosyan}
\affil{Physics Department, University of
Alberta, Edmonton, Canada}

\begin{abstract}
We continue our work on developing techniques for studying turbulence with spectroscopic data.
We show that Doppler-broadened absorption spectral lines, in particularly, saturated absorption
lines,  can be used within the framework of the
earlier-introduced technique termed the Velocity Coordinate spectrum (VCS). The VCS relates the
statistics of fluctuations along the velocity coordinate to the statistics of turbulence, thus it does not
require spatial coverage by sampling directions in the plane of the sky. We consider lines with
different degree of absorption and show that for lines of optical depth less than one, our earlier
treatment of the VCS developed for spectral emission lines is applicable, if the optical depth is
used instead of intensity. This amounts to correlating the logarithms of absorbed intensities. For
larger optical depths and saturated absorption lines, we show, that the amount of information 
that one can use is, inevitably, limited by noise. In practical terms, this means that only wings of the 
line are available for the analysis. In terms of the VCS formalism, this results in introducing an additional  window,
which size decreases with the increase of the optical depth. As a result, strongly saturated absorption lines
carry the information only about the small scale turbulence. Nevertheless, the contrast of the fluctuations
corresponding to the small scale turbulence increases with the increase of the optical depth, which
provides advantages for studying turbulence combining lines with different optical depths. We show that,
eventually, at very large optical depths the Lorentzian profile of the line gets important and extracting information
on velocity turbulence, gets impossible. Combining different absorption lines one can tomography turbulence in
the interstellar gas in all its complexity.
\end{abstract}

\keywords{turbulence -- ISM: general, structure -- MHD -- radio lines: ISM.}

\section{Introduction}

Turbulence is a key element of the dynamics of astrophysical
fluids, including those of interstellar medium (ISM), clusters of galaxies and 
circumstellar regions. The realization of the importance of turbulence
induces sweeping changes, for instance, in the paradigm of ISM. 
It became clear, for instance, that turbulence affects substantially star formation,
mixing of gas, transfer of heat. Observationally it is known that the ISM is turbulent 
on scales ranging from AUs to kpc (see Armstrong
et al 1995, Elmegreen \& Scalo 2004), with an embedded magnetic field
that influences almost all of its properties.

The issue of quantitative descriptors that can characterize turbulence is
not a trivial one (see discussion in Lazarian 1999 and ref. therein). One of the most widely 
used measures is the turbulence
spectrum, which describes the distribution of turbulent fluctuations over
scales. For instance, the famous Kolmogorov model of incompressible turbulence
 predicts that the difference in velocities at
different points in turbulent fluid increases on average
with the separation between points as a cubic root of the separation,
i.e. $|\delta v| \sim l^{1/3}$. In terms of direction-averaged
energy spectrum this gives the famous Kolmogorov
scaling $E(k)\sim 4\pi k^2 P({\bf k})\sim k^{5/3}$, where $P({\bf k})$ 
is a {\it 3D} energy spectrum defined as the Fourier transform of the
correlation function of velocity fluctuations $\xi ({\bf r})=\langle  
\delta v({\bf x})\delta v({\bf x}+{\bf r})\rangle$. Note that in
this paper we use $\langle  ...\rangle$ to denote averaging procedure.

Quantitative measures of turbulence, in particular, turbulence
spectrum, became important recently also due
to advances in the theory of MHD turbulence. As we know, astrophysical
fluids are magnetized, which makes one believe that the correspondence
should exist between astrophysical turbulence and MHD models of the
 phenomenon (see Vazquez-Semadeni et al. 2000, Mac Low \& Klessen 2004,
Bellesteros-Paredes et al. 2007, McKee \& Ostriker 2007 and ref. therein).

 In fact, without observational testing, the application
of theory of MHD turbulence to astrophysics could always be suspect.
Indeed, from the point of view of fluid mechanics astrophysical turbulence 
is characterized by huge Reynolds numbers, $Re$, which is  
the inverse ratio of the
eddy turnover time of a parcel of gas to the time required for viscous
forces to slow it appreciably. For $Re\gg 100$ we expect gas to be
turbulent and this is exactly what we observe in HI (for HI $Re\sim 10^8$).
In fact, very high astrophysical $Re$ and its magnetic counterpart
magnetic Reynolds number $Rm$
 (that can be as high as $Rm\sim 10^{16}$) present a big problem for numerical simulations
that cannot possibly get even close to the astrophysically-motivated numbers. The currently available 3D
simulations can have $Re$ and $Rm$ up to $\sim 10^4$.  Both scale as
the size of the box to the first power, while the computational effort
increases as the fourth power (3 coordinates + time), so the brute force 
approach cannot begin to resolve the controversies related, for example,
to ISM turbulence.

We expect that observational studies of turbulence velocity spectra will 
provide important insights into ISM physics. Even in the case of much more
simple oceanic (essentially incompressible) turbulence, studies of spectra
allowed to identify meaningful energy injection scales. In interstellar,
intra-cluster medium, in addition to that,
 we expect to see variations of the spectral index arising 
from the variations of the degree of compressibility, magnetization, 
interaction of different interstellar phases etc. 

How to get the turbulence spectra from observations is a problem of a long 
standing. While density fluctuations are readily available through
both interstellar scincillations and studies of column density maps,
the more coveted velocity spectra have been difficult to obtain reliably
until very recently. 

Turbulence is associated with fluctuating velocities that cause
fluctuations in the Doppler shifts of emission and absorption
lines. Observations provide integrals of either emissivities or
opacities, both proportional to the local densities, at each velocity
along the line of sight. It is far from trivial to determine the
properties of the underlying turbulence from the observed spectral
line.

Centroids of velocity (Munch 1958) have been an accepted way of 
studying turbulence,  although it was not clear to when and to what extend
the measure really represents the velocity.
Recent studies (Lazarian \& Esquivel 2003, henceforth LE03, Esquivel
\& Lazarian 2005, Ossenkopf et al 2006, Esquivel et al. 2007) have showed that
the centroids are not a good measure for supersonic turbulence, which means
that while the results obtained for HII regions (O'Dell \& Castaneda 1987) are probably OK,
those for molecular clouds are unreliable.        

An important progress in analytical description of the relation between the
spectra of turbulent velocities and the observable spectra of fluctuations
of spectral intensity was obtained in Lazarian \& Pogosyan (2000, henceforth
LP00). This description paved way to two new techniques, which were later
termed Velocity Channel Analysis (VCA) and Velocity Coordinate Spectrum (VCS).

The techniques provide different ways of treating observational data in 
Position-Position-Velocity (PPV) data cubes. While VCA is based on the
analysis of channel maps, which are the velocity slices of PPV cubes,
the VCS analyses fluctuations along the velocity direction. If the slices
have been used earlier for turbulence studies, although the relation between
the spectrum of intensity fluctuations in the channel maps and the underlying
turbulence spectrum was unknown, the analysis of the fluctuations along the
velocity coordinate was initiated by the advent of the VCS theory.

With the VCA  and the VCS one can relate both
 observations and simulations to {\it turbulence theory}. For
 instance, the aforementioned turbulence indexes are very informative,
 e.g. velocity indexes steeper than the Kolmogorov value of $-5/3$
 are likely to reflect formation of shocks, while shallower indexes
 may reflect scale-dependent suppression of cascading (see Beresnyak
 \& Lazarian 2006 and ref. therein).  By associating the variations of
 the index with different regions of ISM, e.g. with high or low star
 formation, one can get an important insight in the fundamental
 properties of ISM turbulence, its origin, evolution and dissipation.

The absorption of the emitted radiation was a concern of the observational
studies of turbulence from the very start of the work in the field 
(see discussion in Munch 1999). A quantitative study of the effects of the absorption was 
performed for the VCA in Lazarian \& Pogosyan (2004, henceforth LP04)
and for the VCS in Lazarian \& Pogosyan (2006, henceforth LP06). In LP06
it was stressed that absorption lines themselves can be used to study 
turbulence. Indeed, the VCS is a unique technique that does not require a
spatial coverage to study fluctuations. Therefore individual point sources
sampling turbulent absorbing medium can be used to get the underlying turbulent
spectra.

However, LP06 discusses only the linear regime of absorption, i.e. when the 
absorption lines are not saturated. This substantially limits the applicability
of the technique. For instance, for many optical and UV absorption lines, e.g.
Mg II, SII, SiII the measured spectra show saturation. This means that a part of 
the wealth of the unique data obtained e.g. by HST and other instruments cannot be handled
with the LP06 technique.

The goal of this paper is to improve this situation. In particular, 
in what follows, we develop a theoretical description that allows to relate the 
fluctuations of the absorption line profiles and the underlying velocity spectra in the saturated regime.

Below, in \S 2 we describe the setting of the problem we address, while our main derivations are in \S 3. The discussion of the 
new technique of turbulence study is provided in \S 4, while the summary is in \S 5. 

\section{Absorption Lines and Mathematical Setting}

While in our earlier publications (LP00, LP04, LP06)
 concentrated on emission line, in
particular radio emission lines, e.g. HI and CO, absorption lines present the researchers with well defined advantages. For instance, they allow
to test turbulence with a pencil beam, suffer less from uncertainties
in path length. In fact, studies of absorption features in the spectra
of stars have proven useful in outlining the gross features of gas kinematics in Milky  way. Recent advances in sensitivity and spectral resolution of spectrographs allow studies of turbulent motions. 

Among the available techniques, VCS is the leading candidate to be used 
with absorption lines. Indeed, it is only with extended sources that the either centroid or VCA studies are possible. At the same time, VCS makes use not of the spatial, but frequency  resolution. Thus, potentially,
turbulence studies are possible if absorption along a single line is available. In reality, information along a few lines of sight, as it shown
in Fig~1 is required to improve the statistical accuracy of the measured spectrum. Using the simulated data sets Chepurnov \& Lazarian (2006ab) 
experimentally established that the acceptable number of lines ranges 
from 5 to 10.

For weak absorption, the absorption and emission lines can be analyzed
in the same way, namely, the way suggested in LP06. For this case,
the statistics to analyse is the squared Fourier transform of the 
Doppler-shifted spectral line, irrespectively of the fact whether this
is an emission or an absorption spectral line. Such a ``spectrum of
spectrum'' is not applicable for saturated spectral lines, which width
is still determined by the Doppler broadening. It is known (see Spitzer 1978) that this
regime corresponds to the optical depth $\tau$ ranging from 10 to $10^3$.
The present paper will concentrate on this regime\footnote{It is known
that for $\tau$ larger than $10^5$ the  line width is determined by
atomic constants and therefore it does not carry information about
turbulence.}.

\begin{figure}[h]
\includegraphics[height=0.6\textwidth]{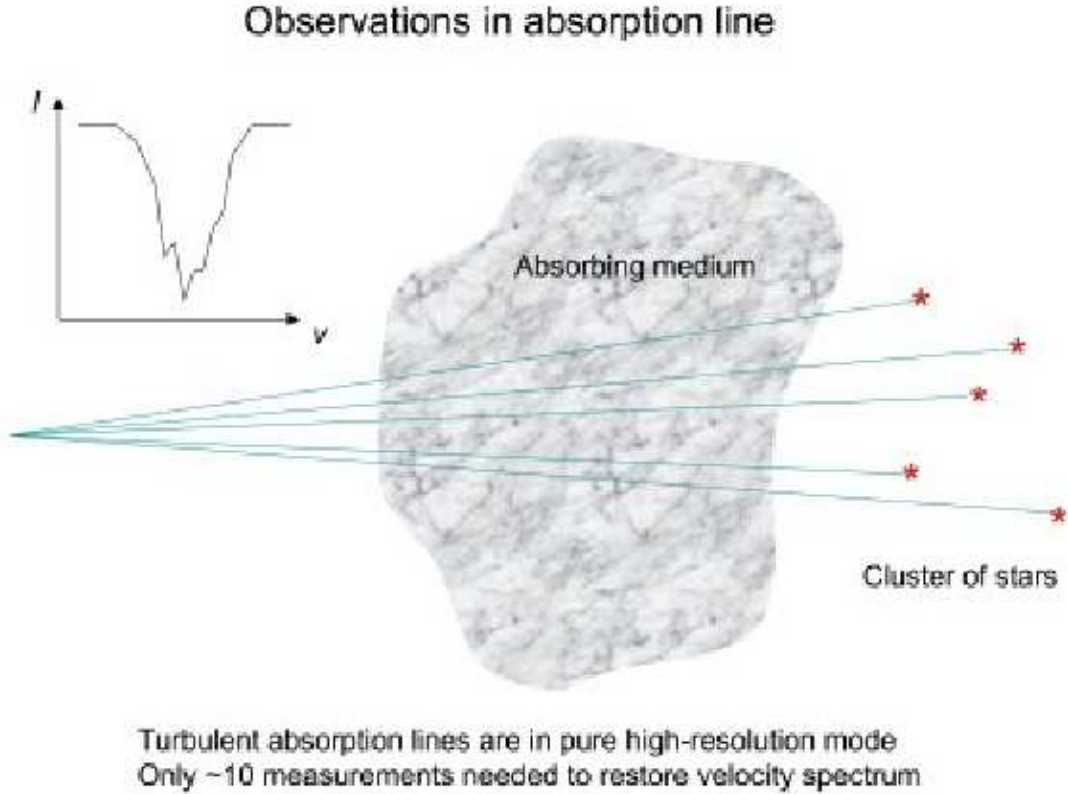}
\caption{Absorption line study of turbulence in a cloud with VCS.}
\end{figure}

Consider the problem in a more formal way.
Intensity of the absorption line at frequency $\nu$ is given as
\begin{equation}
I(\nu,\nu_0) = I_0 e^{-\tau(\nu,\nu_0)}
\end{equation}
where $\tau(\nu)$ is the optical depth. In the limit of vanishing intrinsic 
width of the line $\phi_i(\nu-\nu_0) = \alpha \delta(\nu - \nu_0)$,
the frequency spread of $\tau(\nu,\nu_0)$
is determined solely by the Doppler shift
of the absorption frequency from moving atoms.

The number density of atoms along the line of sight 
moving at required velocity $v \approx \frac{c}{\nu_0} (\nu-\nu_0)$
is
\begin{equation}
\rho_s(v) = \int_0^S dz \rho(z) 
\phi(v-u(z)-v_{reg})
\end{equation}
where $\phi$ is the thermal distribution centered at every point $z$ at
the local mean velocity that is determined by the sum of
turbulent and regular flow at that point. This is the density in PPV coordinate
that we introduced in LP00, so $\tau(\nu)=\alpha(\nu_0) \rho_s(v)$.
The intrinsic line width is accounted for by the convolution
\begin{equation}
\tau(v) =\alpha(\nu_0) \int dw \rho_s(w) \phi_i(w-v)
\end{equation}
or, in more detail,
\begin{equation}
\tau(v) = \alpha(\nu_0) \int_0^S dz n(z) 
\int dw \phi(w-u(z)-v_{gal}(z)) \phi_i(w-v) ~.
\end{equation}
With intrinsic profile given by the Lorenz form
$\phi_i(w-v)=\frac{a/\pi}{(w-v)^2+a^2}$, the inner integral gives the shifted
Voigt profile
\begin{equation}
H(v-u(x)-v_{gal}(x)) = \frac{1}{\pi \sqrt{2 \pi \beta}}
\int dw \frac{a}{(w-v)^2+a^2}
\exp\left(-\frac{(w-u(x)-v_{gal}(x))^2}{2 \beta}\right) ~,
\label{eq:H_detailed}
\end{equation}
so we have another representation
\begin{equation}
\tau(v) = \alpha(\nu_0) \int_0^S dz n(x) H(v-u(x)) ~.
\label{eq:tau_H}
\end{equation}

We clearly see from Eq.~(\ref{eq:tau_H}) that the line is affected both by Doppler shifts\footnote{Speaking formally, one can always make
use of the known atomic constants and get inside into turbulence. This is not practical, however, for an actual spectrum in the presence of
noise.} and atomic constants. 

\section{Fluctuations of Optical Depth}

\subsection{Statistics of Optical Depth Fluctuations}

The optical depth as a function of frequency contains fluctuating component 
arising from turbulent motions and associated density inhomogeneities of the
absorbers. Statistics of optical depth fluctuations along
the line of sight therefore carries information about turbulence in ISM.

The optical depth is determined by the density of the absorbers
in the PPV space, $\rho_s$.  In our previous work we have studied
statistical properties of $\rho_s$ in the context of emission lines,
using both structure function and power spectrum formalisms.
Absorption lines demonstrate several important differences that warrant
separate study.

Firstly, our ability to recover the optical depth from the observed intensity 
\begin{equation}
I(\nu) = I_0 e^{-\tau(\nu)} + N
\end{equation}
depends on the magnitude of the absorption as well as sensitivity of the
instrument and the level of measurement noise $N$.

For lines with low optical depth $\tau_0 < -\ln(N/I_0)$ we can in principle 
measure the optical depth throughout the whole line. 
At higher optical depths, the central part of the line is saturated below
the noise level and the useful information is restricted to the wings of the
line. This is the new regime that is the subject of this paper.

In this regime the data is available over a window of frequencies
limited to velocities high enough so that $\tau(v) < -\ln(N/I_0)$ but not as
high as to have Lorentz tail define the line. Higher the overall optical
depth, narrower are the wings (following Spitzer, at $\tau_0 > 10^5$ the wings
are totally dominated by Lorentz factor). We shall denote this window 
by $W(v-v_0,\Delta)$ where $v_0$ is the velocity that the 
window is centered upon (describing frequency position of the wing)
and $\Delta$ is the wing width. It acts as a mask on the ``underlying'' data
\begin{equation}
\tau(v) \to \tau(v)W(v-v_0,\Delta) ~ ~.
\end{equation}
 
Secondly, fluctuations in the wings of a line are superimposed on the
frequency dependent wing profile. In other words, the statistical properties
of the optical depth are inhomogeneous in this frequency range, with
frequency dependent statistical mean value.
While {\it fluctuations} of the optical depth $\delta
\tau(v)$ that have origin in the turbulence can still be assumed
to be statistically homogeneous,
the mean profile of a wing must be accounted for.

What statistical descriptors one should chose in case of line of sight velocity
data given over limited window? Primary descriptors of a random
field, here $\tau(v)$,
are the ensemble average product of the values of the field at 
separated points --- the two point correlation function
\begin{equation}
\xi_{\tau}(v_1,v_2)=\left\langle \tau(v_1) \tau(v_2) \right\rangle
\end{equation}
and, reciprocally, the average square of the amplitudes of its (Fourier)
harmonics decomposition --- the power spectrum
\begin{equation}
P_{\tau}(k_{v,1} k_{v,2}) = \left\langle \tau(k_{v,1}),\tau^*(k_{v,2}) \right\rangle
\end{equation}
In practice these quantities are measurable if one can replace ensemble average
by averaging over different positions which relies on some homogeneity 
properties of stochastic process. We assume that underlying turbulence is
homogeneous and isotropic. This does not make the optical depth to 
be statistically homogeneous in the wings of the line, but allows to
introduce the fluctuations of $\tau$ on the background of 
the mean profile $\bar \tau(v)$, $\Delta \tau(v)=\tau(v)-\bar \tau(v)$,
which are (LP04) 
\begin{eqnarray}
\xi_{\Delta\tau}(v) &=& 
\left\langle \Delta \tau(v_1) \Delta \tau(v_2) \right\rangle ~,~ ~v=v_1-v_2 \\
P_{\Delta\tau}(k_{v,1} k_{v,2}) &=& 
\left\langle \Delta\tau(k_{v})\Delta\tau^*(k_{v}) \right\rangle
\delta(k_{v_1}-k_{v_2})
\end{eqnarray}
Homogeneous correlation function depends only on a point separation and
amplitudes of distinct Fourier harmonics are independent.  The obvious 
relations are
\begin{eqnarray}
\xi_{\tau}(v_1,v_2) &=& \xi_{\Delta\tau}(v) + \bar \tau(v_1) \bar \tau(v_2) \\
P_{\tau}(k_{v,1} k_{v,2}) &=& P_{\Delta\tau}(k_{v})  
+ \bar \tau(k_{v1}) \bar \tau^*(k_{v2}) 
\end{eqnarray}

Although mathematically the power spectrum is just a Fourier transform of 
the correlation function
\begin{equation}
P_{\Delta \tau}(k_v) = \int e^{i k_v v} \xi_{\Delta\tau}(v) dv ~ ~ ~,
\end{equation}
which of them is best estimated from data  depends on the properties of the signal
and the data.

The power spectrum carries information which is localized to a
particular scale and as such is insensitive to processes that contribute
outside the range of scales of interest, in particular to long-range smooth
variations. On the other hand, determination of Fourier harmonics is
non-local in configuration space and is sensitive to 
specifics of data sampling -- the finite window,
discretization, that all lead to aliasing of power from one scales to another.
The issue is severe if the aliased power is large.

Conversely, the correlation function is localized in configuration space and
can be measured for non-uniformly sampled data. However, at each separation it
contains contribution from all scales and may mix together the physical effects
from different scales. In particular, $\xi(v)$ is not even defined
for power law spectra $P(k_v) \sim k_v^{-n}$ with index $n \ge 1$ (for 
one dimensional data)
\footnote{In physically realistic situations the power law will not, of course, 
extend infinitely to large scales. Mathematical divergence of the correlation
function in practice mean that for steep spectra the largest present scales
the correlation at all, even small separations $v$.}. This limitation is 
relieved if one uses the structure function
\begin{equation}
D_{\Delta\tau}(v)= \left\langle(\Delta\tau(v_1)-\Delta\tau(v_2))^2\right\rangle
\end{equation}
instead, which is well defined for $n < 3$.
The structure function can be thought of as
regularized version of the correlation function
\begin{equation}
D_{\Delta\tau}(v) = 2 \left( \xi_{\Delta\tau}(0) - \xi_{\Delta\tau}(v) \right) 
\end{equation}
that is related to the power spectrum in the same way as the correlation
function, if one excludes the $k_v=0$ mode.

Velocity Coordinate Spectrum studies of LP06 demonstrated that
the expected one dimensional spectrum of PPV density fluctuations 
along velocity coordinate that arise from turbulent motions is
$P(k_v) \sim k_v^{-2/m}$ where $m$ is the index of line-of-sight component
of the velocity structure function.  For Kolmogorov turbulence $m=2/3$ and
 for turbulent motions dominated by shocks $m=1/2$. These spectra
are steep $\sim k_v^{-3}, \sim k_v^{-4}$ which makes the direct measurement of
the structure functions impractical (although for $m>2/3$ the structure function
can be defined). At the same time, in our present studies we deal with
a limited range of data in the wings of the absorption lines, which complicates
the direct measurements of the power spectrum.

Below we first describe the properties of the power spectrum 
$P_{\Delta\tau}(k_v)$ in this case, and next develop the formalism of higher
order structure functions.

\subsection{Power Spectrum of Optical Depth Fluctuations}

Let us derive the power spectrum of the optical depth fluctuations,
$P_\tau(k_v)\equiv\langle \tau(k_v) \tau^*(k_v)\rangle $ Here $k_v$ is wave number
reciprocal to the velocity (frequency) separation between two points on the 
line-of-sight and angular brackets denote an ensemble averaging.
\footnote{note, that due to
having measurements in the finite window, the amplitudes at different
waves numbers are, in general, correlated, $\langle \tau(k_v) 
\tau^*(k_v^\prime)\rangle \ne 0, \quad k_v \ne k_v^\prime $. Here we 
restrict ourselves to diagonal terms only.} 

Fourier transform of the eq.~(\ref{eq:tau_H}) with respect to velocity
is
\begin{equation}
\tau(k_v)= \alpha(\nu_0) \int_0^S dz n(x) 
\int d k_v^\prime 
e^{-|k_v^\prime|a} e^{-\frac{{k_v^\prime}^2 \beta}{2}} 
e^{-i k_v^\prime u(x)}
W(k_v-k_v^\prime, v_0,\Delta)
\label{eq:H_detailed}
\end{equation}
and the power spectrum
\begin{eqnarray}
P_\tau(k_v)&=& \left\langle
\alpha(\nu_0)^2 \int_0^S dz_1 \int_0^S dz_2 n(z_1) n(z_2) \times \right. \\
&&\left. \int d k_v^\prime \int d k_v^{\prime\prime} 
e^{-(|k_v^\prime|+|k_v^{\prime\prime}|)a} 
e^{-\frac{\left({k_v^\prime}^2 + {k_v^{\prime\prime}}^2\right) \beta}{2}} 
e^{-i [k_v^\prime u(z_1)-k_v^{\prime\prime} u(z_2)]}
W(k_v-k_v^\prime) W^*(k_v-k_v^{\prime\prime})
\right\rangle \nonumber
\end{eqnarray}
which is useful to express using average velocity
$u_+ = (u(z_1)+u(z_2))/2$ and velocity
difference $u=u(z_1)-u(z_2)$, as well as correspondent
 variables for the wave numbers
$k_v^+=(k_v^\prime+k_v^{\prime\prime})/2$ and
$k_v^-=k_v^\prime-k_v^{\prime\prime}$, as
\begin{eqnarray}
P_\tau(k_v)&=& \left\langle
\alpha(\nu_0)^2 \int_0^S dz_1 \int_0^S dz_2  n(z_1) n(z_2)
\int_{-\infty}^\infty d k_v^+ 
e^{-{k_v^+}^2 \beta} e^{i k_v^+ u} \right.  \label{eq:Ptau_gen}\\
&\times&\left. \int_{-\infty}^\infty
 d k_v^- e^{-\frac{1}{4}{k_v^-}^2 \beta} e^{i k_v^- u_+}
e^{-(|k_v^\prime|+|k_v^{\prime\prime}|)a}
W\left(k_v-k_v^+-\frac{k_v^-}{2}\right)
W^*\left(k_v-k_v^++\frac{k_v^-}{2}\right)
\right\rangle \nonumber
\end{eqnarray}
The fluctuating, random quantities, over which the averaging is performed
are the density $n(z)$ and the line-of-sight component of the velocity of the
absorbers $u(z)$, varying along the line of sight . In our earlier papers (see LP00, LP04) we argued that
in many important cases they can be considered as uncorrelated between
themselves, so that
\begin{equation}
\left\langle n(z_1) n(z_2) e^{i k_v^+ u} e^{i k_v^- u_+} \right\rangle
=  \xi(|z_1-z_2|) e^{-\frac{1}{2}{k_v^+}^2 D_z(|z_1-z_2|)}
e^{-\frac{1}{4}{k_v^-}^2 \left[D_z(S)-D_z(|z_1-z_2|)/2\right]} ~ ~,
\label{eq:maxprof_average}
\end{equation} 
where $\xi(|z_1-z_2|)$ is the correlation function of the density
of the absorbers and $D_z(|z_1-z_2|)$ is the structure function of 
their line-of-sight velocity due to turbulent motions.
 $D_z$ is expected to saturate at the value
$D_s(S)$ for separations of the size of the absorbing cloud $S$.
The dependence of $\xi(z)$ and $D_z(z)$ only on spatial separation between
a pair of absorbers reflects the assumed statistical homogeneity of
the turbulence model. Introducing $z=z_1-z_2$ and performing integration
over $z_+=(z_1+z_2)/2$ one obtains \footnote{Absolute values in the Fourier
image of the Lorentz transform require separate consideration of each quadrant
of wave numbers. This is carried out in the Appendix.}
\begin{eqnarray}
P_\tau(k_v)&=& \alpha^2 S \int_0^S \!\!\!\! dz (S-|z|) \xi(z) 
\int_0^\infty \!\!\!\! d k_v^+ e^{-2k_v^+a}
e^{-\frac{1}{2}{k_v^+}^2 D^-}
\int_0^{\infty} \!\!\!\! d k_v^- 
e^{-\frac{1}{4}{k_v^-}^2 D^+}
\tilde W \left(k_v,k_v^+,\frac{k_v^-}{2}\right)
\nonumber \\
&+& correction~terms~(see~Appendix) 
\end{eqnarray}
where $D^- = D_z(|z_1-z_2|) + 2 \beta$, $D^+=D_z(S)-D_z(|z_1-z_2|)/2+\beta$
and symmetrized window $\tilde W$ is defined in the Appendix. 

If one has the whole line available for analysis, the masking window 
will be flat with $\delta(k)$-function like Fourier transform. The combination
of the windows in the power spectrum will translate to  
$\delta(k_v-k_v^+) \delta(k_v^-)$ and 
\begin{equation}
P_\tau(k_v) \sim 2 \alpha(\nu_0)^2 S 
e^{-2k_v a-{k_v}^2 \beta}
\int_0^S dz (S-|z|) \xi(z) e^{-\frac{1}{2}{k_v}^2 D^-}
\end{equation}
Masking the data has the effect of aliasing modes of the large scales that
exceed the available data range, to shorter wavelength. This is represented
by the convolution with Fourier image of the mask. Secondary effect is
the contribution of the modes with different wave numbers $k_v^- \ne 0$ to
the diagonal part of the power spectrum. This again reflects the situation 
that different Fourier components are correlated in the presence of the mask.

\begin{figure}[h]
\includegraphics[height=.3\textwidth]{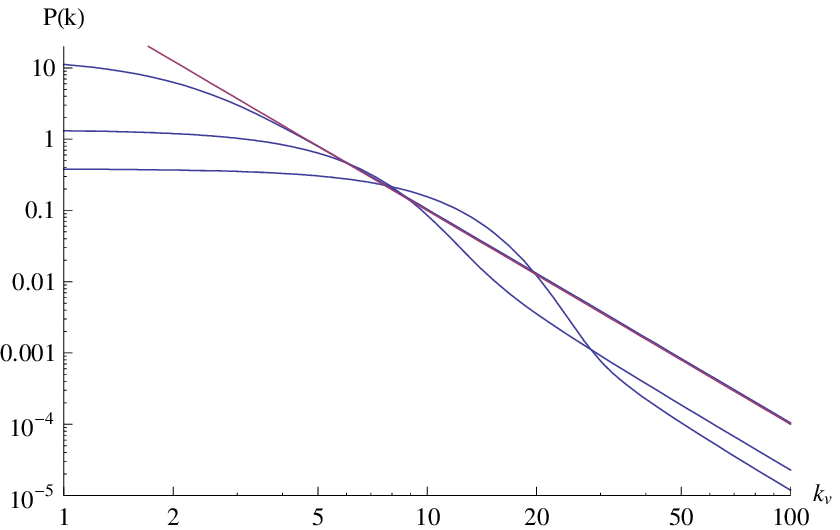}
\includegraphics[height=.3\textwidth]{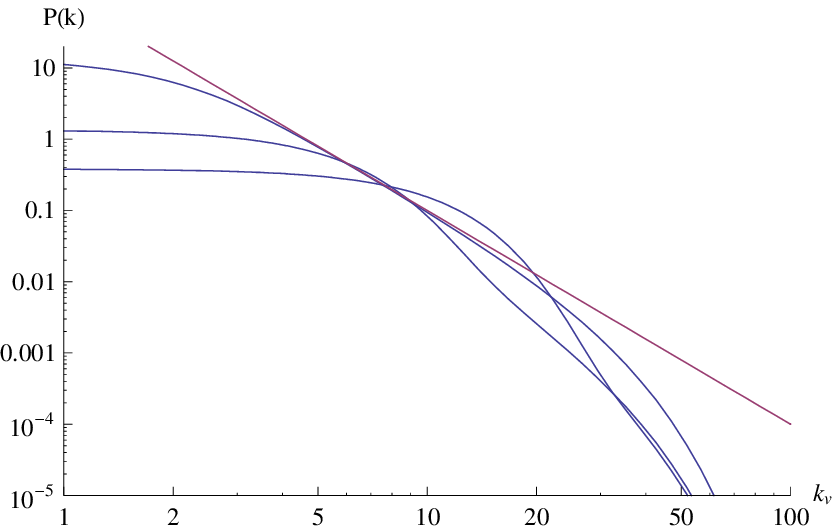}
\caption{Power spectrum of the optical depth fluctuations from line wings.
All parameters, $k_v$, $\Delta$, $\sqrt{\beta}$, $v_1$ are dimensionless,
in the units of $\sqrt{D_z(S)}$, the variance of the turbulent velocity at
the scale of the cloud. Intrinsic line broadening is neglected. 
Only the effect of the turbulent motions and not spatial inhomogeneity of the
absorbers is taken into account. The underlying scaling of the turbulent
velocities is Kolmogorov, $m=2/3$.
The left panel illustrates the power aliasing due to finite width of the window.
The power spectrum is plotted, from top to bottom, for $\Delta=1,0.2.0.1$, i.e
the widths of the wing ranges from the complete line to one-tenth of the
line width. The straight line shows the power law 
$P(k_v) \propto k_v^{-2/m} = k_v^{-3}$ expected under ideal observational 
circumstances.
One finds that for the ideal Gaussian mask the underlying spectrum is 
recovered for $k_v > 3 \Delta^{-1}$. The right panel shows the
modification of the spectrum due to thermal broadening, which is
taken at the level $\sqrt{\beta}=\frac{1}{30} \sqrt{D_z(S)}$. Thermal
effects must be accounted for for $k_v > 1/(3 \sqrt{\beta})$.}
\label{fig:spectrum}
\end{figure}

To illustrate the effects of the mask, let us assume that we select
the line wing with the help of a Gaussian mask centered in the middle of the wing at $v_1$
\begin{eqnarray}
W(v-v_1,\Delta) &=& 
e^{-\frac{(v-v_1)^2}{2 \Delta^2}} \\
W(k_v) &=& \sqrt{2 \pi} \Delta e^{-i k_v v_1} e^{-\frac{1}{2} k_v^2 \Delta^2}
\end{eqnarray}
that gives
\begin{equation}
\tilde W = 
4 \pi \Delta^2 \cos\left(\frac{k_v^-}{2} v_1 \right)
e^{-\frac{{k_v^-}^2 \Delta^2} {4}}
\left(e^{-(k_v-k_v^+)^2\Delta^2}+e^{-(k_v+k_v^+)^2 \Delta^2} \right)
\end{equation}
All integrals can then be carried out to obtain
\begin{eqnarray}
P(k_v) &\sim& \alpha^2 S 
\int_0^S dz (S-z) \xi(z) 
\frac{\Delta^2 \exp\left[-\frac{1}{4}\frac{v_1^2}{D^+ + \Delta^2}\right]}
{\sqrt{(D^+ + 2 \Delta^2)(D^- + 2 \Delta^2)}}
\exp\left[-\frac{\Delta^2 D^-}{D^- + 2 \Delta^2} k_v^2\right] 
\\
&& \times
\exp\left[\frac{2 a^2}{D^-+2 \Delta^2}\right]
\left\{ \exp\left[\frac{-4 a \Delta^2 k_v}{D^- +2\Delta^2}\right] 
 \mathrm{Erfc}
\left[\frac{\sqrt{2}(a-\Delta^2 k_v)}{\sqrt{D^-+2\Delta^2}}
\right] + (k_v \to -k_v) \right\} \nonumber
\end{eqnarray}
The following limits (taking $k_v > 0$) are notable:
\begin{eqnarray}
\Delta \to \infty &:& \\
P(k_v) &\propto& \alpha(\nu_0)^2 S 
\int_0^S dz (S-z) \xi(z) \exp\left[-\frac{1}{2} k_v^2 D^- -2 a k_v\right]
\nonumber \\
a \to 0 &:& \\
P(k_v) &\propto& \alpha(\nu_0)^2 S 
\int_0^S dz (S-z) \xi(z)
\frac{\Delta^2 \exp\left[-\frac{1}{4}\frac{v_1^2}{D^+ + \Delta^2}\right]}
{\sqrt{(D^+ + 2 \Delta^2)(D^- + 2 \Delta^2)}}
\exp\left[-\frac{\Delta^2 D^-}{D^- + 2 \Delta^2} k_v^2\right] \nonumber
\end{eqnarray}
The last expression particularly clearly demonstrates the effect of the window,
which width in case of the line wing is necessarily necessarily
limited by $\Delta^2 < D(S)+2 \beta$.
The power spectrum is corrupted at scales $k_v < \Delta^{-1}$, but
still maintains information about turbulence statistics for 
$k_v \gg \Delta^{-1}$. Indeed, in our integral representation the
power spectrum at $k_v$ is determined by the linear scales such that
$\frac{\Delta^2 D^-}{D^- + 2 \Delta^2} k_v^2 < 1$ which translates into
$D^- < 2 \Delta^2/(k_v^2 \Delta^2 -1)$. Thus, if $k_v \Delta \gg 1$
over all scales defining power at $k_v$ one has $\Delta \gg D^-$ and
there is no significant power aliasing.\footnote{In this regime the first
factor is essentially constant since $D^+(z)=D_z(S)+2 \beta - D^-(z)/2$ varies
little, $D^+(z) \approx D^+(0)$ in the interval 
$2 \beta < D^- < 2 k_v^{-2} \ll 2 \Delta^2 < D(S) + 2 \beta$.}
For intermediate scales there is
a power aliasing as numerical results demonstrate in Figure~\ref{fig:spectrum}.
Doppler broadening described by $D^-$ incorporates both turbulent and thermal 
effects. Thermal effects are especially important in case of narrow line wings,
since the range of the wavenumbers relatively unaffected by both thermal motions
and the mask is limited $1/sqrt{\beta} > k_v > \Delta^{-1}$ and exists only
for relatively wide wings $\Delta > \sqrt{\beta}$. For narrower wings the 
combined turbulent and thermal profile must be fitted to the data, possibly 
determining the temperature of the absorbers at the same time. 
This recipe is limited by the assumption that the temperature of the gas is
relatively constant for the absorbers of a given type.

We should note that the Gaussian window provides one of the ideal cases,
limiting the extend of power aliasing since the window Fourier image falls off
quickly.
One of the worst scenarios is represented by sharp top hat mask, which
Fourier image falls off only as $k_v^{-2}$ spreading the power from 
large scales further into short scales. For steep spectra that we have in
VCS studies all scales may experience some aliasing. This argues for extra care
while treating the line wings through power spectrum or for use of alternative
approaches.

\subsection{Second order structure function}
Second order structure function provides an alternative to power spectrum
measurement in case of steep spectra with the data limited to the section
of the lines. 
The second order structure function of the fluctuations
of the optical depth can be defined as
\begin{equation}
dD_{\Delta\tau}(v) = \frac{1}{2} 
\left\langle \left(\Delta\tau(v_1+v)+\Delta\tau(v_1-v)-2 \Delta\tau(v_1)
\right)^2 \right\rangle
\end{equation}
It represents additional regularization of the correlation function beyond
the ordinary structure function
\begin{eqnarray}
dD_{\Delta\tau}(v) &=& 3 \xi_{\Delta\tau}(0) - 4 \xi_{\Delta\tau}(v) 
+ \xi_{\Delta\tau}(2v) \\
&=& 2 D_{\Delta\tau}(v) - \frac{1}{2} D_{\Delta\tau}(2 v)
\end{eqnarray}

$D_{\Delta\tau}(v)$ is proportional to three dimensional velocity space density
structure function at zero angular separations $ d_\rho(0,v) $,
discussed in LP06. Using the results of LP06 for $ d_\rho(0,v) $ we obtain
\begin{eqnarray}
dD_{\Delta\tau}(v) &\sim&
\left(\frac{r_0}{S}\right)^\gamma 
\int_{-1}^1 {\mathrm d}\hat z \;
\frac{1}{|\hat z|^{\gamma+m/2}} \left[ 3 -
4 \exp\left(-\frac{{\hat v}^2}{2 |\hat z|^m}\right)
+ \exp\left(-\frac{2 {\hat v}^2}{|\hat z|^m}\right)
\right] \nonumber \\
&\propto& \frac{{\bar \rho}^2 S^2}{ D_z(S)} \frac{1}{m}
\left(\frac{r_0}{S}\right)^\gamma 
\left[ \hat v^{2p} \Gamma(-p) \left(2^{p-1}-2^{1-p}\right)
+ \frac{2^{4 p-6}}{p-2} \hat v^4 + O( \hat v^6) \right]
\label{eq:d_vV}
\end{eqnarray}
where $p=(1-\gamma)/m-1/2 > 0$ and $\gamma$ is the correlation index
that describes spatial inhomogeneities of the absorbers.
To shorten intermediate formulas, the dimensionless quantities
$\hat v=v/D_z^{1/2}(S)$, $\hat z=z/S$, $\hat r = r/S$ are introduced.
The first term in the expansion contains information about the underlying
field, while the power law series represent the effect of boundary conditions
at the cloud scale. In contrast to ordinary correlation function
they are not dominant until $p \ge 2$, i.e for $(1-\gamma)/m < 5/2$ 
the second order structure function is well defined. When turbulent motions
provide the dominant contribution to optical depth fluctuations, $\gamma=0$,
we see that measuring the $dD_\tau$ one can recover the turbulence scaling 
index if $m  > 2/5$, which includes both interesting cases of Kolmogorov
turbulence and shock-dominated motions. This condition is replaced
by $m > 2/5(1-\gamma)$ if the density fluctuations, described by correlation
index $\gamma$, are dominant.

At sufficiently small scales the second order structure function has the same
scaling as the first order one
\begin{equation}
dD_{\Delta\tau}(v) \propto
\frac{{\bar \rho}^2 S^2}{ D_z(S)} \frac{1}{m}
\left(\frac{r_0}{S}\right)^\gamma 
\left(2^{p-1}-2^{1-p}\right) \Gamma(-p) ~ .
\; \hat v^{2p}
\end{equation}

A practical issue of measuring the structure functions directly in the wing of
the line is to take into account the line profile. The directly accessible
\begin{equation}
dD_{\tau}(v_1,v) = \frac{1}{2} 
\left\langle \left(\tau(v_1+v)+\tau(v_1-v)-2 \tau(v_1) \right)^2 \right\rangle
\end{equation}
is related to the structure function of the fluctuations as
\begin{equation}
dD_{\Delta\tau}(v) =  dD_{\tau}(v_1,v) 
-\frac{1}{2}\left(\bar \tau(v_1+v)+\bar \tau(v_1-v) - 2 \bar \tau(v_1)\right)^2
\end{equation}
where the mean profile of the optical depth $\bar \tau$ is related to the
mean profile of PPV density $\bar \rho_s$ given in the Appendix B of LP06.
At small separations $v$, the correction to the structure function due to
mean profile behaves as $v^4$ and is subdominant. 

The price one pays when utilizing higher-level structure function is their
higher sensitivity to the noise in the data. While correlation function itself
is not biased by the noise except at at zero separations (assuming noise is
uncorrelated
\begin{equation}
\left\langle [\tau(v_1)+ N(v_1)][\tau(v_1+v)+N(v_1+v)]\right\rangle = 
\xi_\tau(v) + \langle N^2 \rangle \delta(v)
\end{equation}
already the structure function is biased by the noise which contributes to
all separations
\begin{equation}
\left\langle [\tau(v_1)+ N(v_1)-\tau(v_1+v)-N(v_1+v)]^2\right\rangle = 
D_\tau(v) + 2  \langle N^2 \rangle 
\end{equation}
This effect is further amplified for the second order structure function
\begin{equation}
\frac{1}{2}
\left\langle \left[\tau(v_1+v)+\tau(v_1-v)-2 \tau(v_1)
+N(v_1+v)+N(v_1-v)-2N(v_1)\right]^2 \right\rangle
= dD_\tau(v) + 3 \langle N^2 \rangle 
\end{equation}
The error in the determination of the structures function of higher order
due to noise also increases.

\subsection{Comparison of the approaches}

Structure functions and power spectra are used interchangeably in the theory of turbulence (see Monin \& Yaglom 1975). However, complications arise when spectra are "extremely steep", i.e. the corresponding structure function of fluctuation grows as $x^\aleph$, $\aleph>2$. For such
random fields, one cannot use ordinary structure functions, while the one dimensional Fourier transforms that is employed in VCS corresponds to the power spectrum of $k_v^{-1-\aleph}$ is well defined. 

As a rule, one does not have to deal with so steep spectra in theory of turbulence (see,
however, Cho et al. 2002 and Cho \& Lazarian 2004). Within the VCS, such "extremely steep" spectra emerge naturally, even when the
turbulence is close to being Kolmogorov.  This was noted in LP06, where the spectral approach was presented as the correct one to
studying turbulence using fluctuations of intensity along v-coordinate. 

The disadvantage of the spectral approach is when the data is being limited by a non-Gaussian window function. Then the contributions from the scales determined by the window function may interfere in the obtained spectrum at large $k_v$. An introduction of an additional more narrow
Gaussian window function may mitigate the effect, but limits the range of $k_v$ for which turbulence can be studied. Thus, higher order structure
functions (see the subsection above), is advantageous for the practical data handling. 

In terms of the VCA theory, we used mostly spectral description in LP00, while in LP04, dealing with absorption, we found advantageous to 
deal with real rather than Fourier space. In doing so, however, we faced the steepness of the spectrum along the v-coordinate and provided a
transition to the Fourier description to avoid the problems with the "extremely steep" spectrum. Naturally, our approach of higher order structure
functions is applicable to dealing with the absorption within the VCA technique.

\section{Discussion}

\subsection{Simplifying assumptions}

In the paper above we have discussed the application of VCS to strong 
absorption lines. The following assumption were used. First of all, considering
the radiative transfer we
neglected the effects of 
stimulated emission. This assumption is well satisfied for
optical or UV absorption lines (see Spitzer 1979). Then, we assumed that the
radiation is coming from a point source, which is an excellent approximation
for the absorption of light of a star or a quasar. Moreover, we disregarded
the variations of temperature in the medium.

Within our approach the last assumption may be most questionable. Indeed,
it is known that the variations of temperature do affect absorption lines. 
Nevertheless, our present study, as well as our earlier studies, prove that
the effects of the variations of density are limited. It is easy to see that
the temperature variations can be combined together with the density ones
to get effective renormalized ``density'' which effects we have already
quantified.

Our formalism can also be generalized to include a more sophisticated radiative
transfer and the spatial  extend of the radiation source. In the latter case
we shall have to consider both the case of a narrow and a broad telescope beam,
the way it has been done in LP06. Naturally, the expressions in LP06 for a broad
beam observations can be straightforwardly applied to the absorption lines, substituting 
the optical depth variations instead of intensities. The advantage of the extended source is that not only VCS, 
but also VCA can be used (see Deshpande et al. 2000). As a disadvantage of an extended
source is the steepening of the observed v-coordinate spectrum for studies of unresolved turbulence.
This, for instance, may require employing even higher order structure functions, if one has to
deal with windows arising from saturation of the absorption line.

In LP06 we have studied the VCS technique in the presence of absorption and
formulated the criterion for the fluctuations of intensity to reliably reflect
the fluctuations in turbulent velocities. In this paper, however, we used
the logarithms of intensities and showed that this allows turbulence studies
beyond the regime, at which fluctuations of intensity would be useful. 

A similar approach, potentially, is applicable to the VCS for studies of
emitted radiation in the presence of significant absorption. Indeed, the
equation of the radiative transfer provides us in this case with
\begin{equation}
I_{\bf X}(v)=
\frac{\epsilon}{\alpha}\left[1-{\mathrm e}^{-\alpha \rho_s({\bf X},v)}\right]~~~,
\label{simplified}
\end{equation}
where the first term is a constant, that can be subtracted. If this is done,
the logarithm can be taken of the intensity. Potentially, this should allow
VCS studies of emission lines where, otherwise, absorption distorts the 
statistics.

The difficulty of such an approach is the uncertainty of the base level
of the signal. Taking logarithm is a non-linear operation that may 
distort the result, if the base level of signal is not accounted for properly. However, the advantage of the approach that potentially it allows
studies of velocity turbulence, when the traditional VCA and VCS fail. 
Further research should clarify the utility of this approach.

\subsection{Prospects of Studying ISM Turbulence with Absorption Lines}

The study of turbulence using the modified VCS technique above should be reliable for optical depth $\tau$ up to $10^3$. For this range of
optical depth, the line width is determined by Doppler shifts rather
 than the atomic constants. While formally the entire line profile 
provides information about the turbulence, in reality, the flat saturated part of the profile will contain only noise and will not be useful
for any statistical study. Thus, the wings of the lines will contain
signal.

As several absorption lines can be available along the same line of
sight, this allows to extend the reliability of measurements combining
them together. We believe that piecewise analyses of the wings belonging
to different absorption lines is advantageous.

The actual data analysis may employ fitting the data with models, that, apart from the spectral index,
specify the turbulence injection scale and velocity dispersion, as this is done in Chepurnov et al. (2006).

Note, that measurements of turbulence in the same volume using different absorption lines can provide 
complementary information. Formally, if lines with weak absorption,
i.e. $\tau_0<1$ are available, there is no need for other measurements.
However, in the presence of inevitable noise, the situation may be far from
trivial. Naturally, noise of a constant
level, e.g. instrumental noise, will affect more weak absorption lines. 
The strong absorption lines, in terms of VCS sample turbulence only for
sufficiently large $k_v$. This limits the range of turbulent scales that can
be sampled with the technique. However, the contrast that is obtained with the
strong absorption lines is higher, which provides an opportunity of increasing
signal to noise ratio for the range of $k_v$ that is sampled by the absorption
lines. If, however, a single strong  absorption line is used, an analogy with
a two dish radio interferometer is appropriate. Every dish of the
radio interferometer  samples spatial frequencies in the range approximately
$[1/\lambda, 1/d]$, where $\lambda$ is the operational wavelength,
$d$ is the diameter of the dish. In addition, the radio interferometer samples
the spatial frequency $1/D$, where $D$ is the distance between the dishes.
Similarly, a strong absorption line provides with the information on turbulent
velocity at the largest spatial scale of the emitting objects, as well as the
fluctuation corresponding to the scales $k_{window}, k_{abs}$.

In LP06 we concentrated on obtaining asymptotic regimes for studying turbulence. 
At the same time in Chepurnov et al. (2006) fitting models of turbulence to the data 
was attempted. In the latter approach non-power law observed spectra can be used, which
is advantageous for actual data, for which the range of scales in $k_v$ is rather limited. Indeed,
for HI with the injection velocities of 10 km/s and the thermal velocities of 1 km/s provides an order  of
magnitude of effective
"inertial range". Correcting for thermal velocities one can increase this range by a factor, which depends on the
signal  to noise ratio of the data. Using heavier species rather than hydrogen one can increase
the range by a factor $(m_{heavy}/m_{H})^{1/2}$. This may or may not be enough for observing
good asymptotics. 

We have seen in \ref{} that for absorption lines the introduction of windows determined by the width of the line wings
introduces additional distortions of the power spectrum. However, this is not a problem if, instead of asymptotics,
fitting of the model is used. Compared to the models used in Chepurnov et al. (2006) the models for
absorption lines should also have to model the window induced by the absorption. The advantage is, however,
that absorption lines provide a pure pencil beam observations.

\subsection{Comparison and synergy with other techniques}

Formally, there exists an extensive list of
different tools to study turbulence that predated our studies (see Lazarian 1999 and ref. therein).
However, a closer examination shows that this list is not as impressive as it looks.
Moreover, our research showed that some techniques may provide confusing, 
if not erroneous, output, unless theoretical understanding of what they measure is
achieved. For instance, we mentioned in the introduction an example of the erroneous 
application of velocity centroids to supersonic molecular cloud data. Note, that clumps
and shell finding algorithms would find a hierarchy of clumps/shells for synthetic 
observations obtained with {\it incompressible} simulations. This calls for a more cautious
approach to the interpretation of the results of some of the accepted techniques.

For instance, the use of different wavelets for the analysis of data is
frequently treated in the literature as different statistical
techniques of turbulence studies (Gill \& Henriksen 1990, Stutzki et al. 1998, Cambresy 1999, Khalil et al. 2006), which creates an
illusion of an excessive wealth of tools and approaches. In reality,
while Fourier transforms use harmonics of $e^{i{\bf kr}}$, wavelets use
more sophisticated basis functions, which may be more appropriate for
problems at hand.  In our studies we also use wavelets both to analyze
the results of computations (see Kowal \& Lazarian 2006a) and synthetic
maps (Ossenkopf et al. 2006, Esquivel et al. 2007), along with or
instead of Fourier transforms or correlation functions. Wavelets may
reduce the noise arising from inhomogeneity of data, but we found in
the situations when correlation functions of centroids that we studied
were failing as the Mach number was increasing, a popular wavelet ($\Delta$-variance) was also failing
(cp. Esquivel \& Lazarian 2005, Ossenkopf et al. 2006, Esquivel et
al. 2007).  

While in wavelets the basis functions are fixed, a more
sophisticated technique, Principal Component Analysis (PCA), chooses
basis functions that are, in some sense, the most descriptive.
Nevertheless, the empirical relations obtained with PCA for extracting
velocity statistics provide, according to Padoan et al. (2006), an
uncertainty of the velocity spectral index of the order $0.5$ (see also Brunt et al. 2003), which is too
large for testing most of the turbulence theories. In addition,
while our research in LP00 shows that for density spectra $E_{\rho}\sim
k^{-\alpha}$, for $\alpha<1$ both velocity and density fluctuations
influence the statistics of PPV cubes,  no dependencies of PPV
statistics on density have been reported so far in PCA studies. This also
may reflect the problem of finding the underlying relations
empirically with data cubes of limited resolution. The latter provides
a special kind of shot noise, which is discussed in a number of papers
(Lazarian et al. 2001, Esquivel et al. 2003, Chepurnov \& Lazarian 2006a).

 {\it Spectral Correlation Function (SCF)}
 (see Rosolowsky et al. 1999 for its original form) is another way to study turbulence. Further
 development of the SCF technique in Padoan et al. (2001) removed the
 adjustable parameters from the original expression for the SCF and
 made the technique rather similar to VCA in terms of the
 observational data analysis. Indeed, both SCF and VCA measure
 correlations of intensity in PPV ``slices'' (channel maps with a
 given velocity window $\Delta v$), but if SCF treats
 the outcome empirically, the analytical relations in Lazarian \&
 Pogosyan (2000) relate the VCA measures to the underlying velocity
 and density statistics. Mathematically, SCF contains additional square roots and
 normalizations compared to the VCA expressions. Those make the analytical
 treatment, which is possible for simpler VCA expressions, prohibitive. One might
 speculate that, similar to the case of conventional centroids and not normalized
 centroids introduced in Lazarian \& Esquivel (2003), the actual difference between
 the statistics measured by the VCA and SCF is not significant. 
 
 In fact, we predicted several physically-motivated
 regimes for VCA studies. For instance, slices are ``thick'' for
 eddies with velocity ranges less than $\Delta v$ and ``thin''
 otherwise.  VCA relates the spectral index of intensity fluctuations
 within channel maps to the thickness of the velocity channel and to
 the underlying velocity and density in the emitting turbulent
 volume\footnote{We showed that much of the earlier confusion stemmed from
 different observational groups having used velocity channels of
 different thicknesses (compare, e.g. , Green 1993 and Stanimirovic et al. 1999).}.
 In the VCA these variations of indexes with
 the thickness of PPV ``slice'' are used to disentangle velocity and
 density contributions. We suspect that similar ``thick" and ``thin"
 slice regimes should be present in the SCF analysis of data, but they
 have not been reported yet.  While the VCA can be used for all the
 purposes the SCF is used (e.g. for an empirical comparisons of
 simulations and observations), the opposite is not true. In fact,
 Padoan et al. (2004) stressed that VCA eliminates errors inevitable
 for empirical attempts to calibrate PPV fluctuations in terms of the
 underlying 3D velocity spectrum. 
 
 {\it VCS} is a statistical tool that uses the information of fluctuations along the
 velocity axis of the PPV. Among all the tools that use spectral data, including the VCA, it is unique, as it {\it does not} 
 require spatial resolution. This is why, dealing with the absorption lines, where good spatial coverage
 is problematic, we employed the VCS. Potentially, having many sources sampling the object one can create PPV cubes
 and also apply the VCA technique. However, this requires very extended data sets, while for the VCS sampling with 5 or 10 
 sources can be sufficient for obtaining good statistics (Chepurnov \& Lazarian 2006a). 

We feel that dealing with the ISM turbulence, it is synergetic to combine different approaches. For the wavelets
used their relation with the underlying Fourier spectrum is usually well defined. Therefore the formulation
of the  theory (presented in this work, as well as, in our earlier papers in terms of the Fourier transforms) in
terms of wavelets is straightforward. At the same time, the analysis
of data with the wavelets may be advantageous, especially, in the situations when one has to deal with
window functions.

\section{Summary}

In the paper above we have shown that 
\begin{itemize}
\item Studies of turbulence with absorption lines are possible with the VCS technique if, instead of intensity $I(\nu)$, one uses
the logarithm of the absorbed intensity $\log I_{abs}(\nu)$, which is equivalent to the optical depth $\tau (\nu)$.

\item In the weak absorption regime, i.e. when the optical depth at the middle of the absorption line is less than unity, the analysis
of the  $\tau(\nu)$ coincides with the analysis of intensities of emission for ideal resolution that we discussed in LP06.

\item In the intermediate absorption retime, i.e. when the optical depth at the middle of the absorption line is  larger than unity, but
less than $10^3$, the wings of the absorption line can be used for the analysis. The saturated part of the line is expected
to be noise dominated.

\item The higher the absorption, the less the portion of the spectrum corresponds to the wings available for the analysis. In terms of
the mathematical setting this introduces and additional window in the expressions for the VCS analysis. However, the contrast
of the small scale fluctuations increases  with the decrease of the window.

\item For strong absorption regime, the broadening is determined by Lorentzian wings of the line and therefore no  information
on turbulence is available.
 
\end{itemize}

\appendix  
\section{Derivation of the Power spectrum $P_\tau$}
\label{App:A}

Following eqns.~(\ref{eq:Ptau_gen},\ref{eq:maxprof_average}) the power spectrum of the optical depth
 is
\begin{eqnarray}
P_\tau(k_v)&=&
\alpha(\nu_0)^2 \int_0^S \!\! dz_1 \int_0^S \!\! dz_2 \; \xi(|z_1-z_2|)
\int_{-\infty}^\infty d k_v^\prime  \int_{-\infty}^\infty \!\! d k_v^{\prime\prime}  \\
&\times& e^{-\frac{1}{2}{k_v^+}^2 D^-}
 e^{-\frac{1}{4}{k_v^-}^2 D^+} e^{-(|k_v^\prime|+|k_v^{\prime\prime}|)a}
W\left(k_v-k_v^+-\frac{k_v^-}{2}\right)
W^*\left(k_v-k_v^++\frac{k_v^-}{2}\right)
\nonumber
\end{eqnarray}
where $D^- = D_z(|z_1-z_2|) + 2 \beta$ and $D^+=D_z(S)-D_z(|z_1-z_2|)/2+\beta$ while
$k_v^+=(k_v^\prime+k_v^{\prime\prime})/2$ and $ k_v^-=k_v^\prime-k_v^{\prime\prime}$.
Since the mask is real, $W^*(k_v)=W(-k_v)$.

To deal with absolute values in the Lorentz transform, we split integration regions in quadrants
I -- $(k_v^\prime > 0 , k_v^{\prime\prime} > 0)$,  II -- $(k_v^\prime < 0 , k_v^{\prime\prime} > 0)$,
III -- $(k_v^\prime  <  0 , k_v^{\prime\prime} < 0)$ and IV -- $(k_v^\prime > 0 , k_v^{\prime\prime} < 0)$.
Integration over quadrants III  and IV can be folded into integration over  regions I and II respectively by substitution $k_v^\prime \to - k_v^\prime,  k_v^{\prime\prime} \to - k_v^{\prime\prime} , k_v^+ \to -k_v^+,
k_v^- \to -k_v^-$.  Writing out only integration over $k_v$
\begin{eqnarray}
I+III &:& 
\int_{0}^\infty d k_v^\prime  \int_{0}^\infty \!\! d k_v^{\prime\prime}  
e^{-\frac{1}{2}{k_v^+}^2 D^-}
 e^{-\frac{1}{4}{k_v^-}^2 D^+} e^{-2 a k_v^+}  \\
 && \times
\left[ W\left(k_v-k_v^+-\frac{k_v^-}{2}\right)
W\left(k_v^+ -k_v -\frac{k_v^-}{2}\right) +
W\left(k_v + k_v^+ +\frac{k_v^-}{2}\right)
W\left(-k_v - k_v^+ +\frac{k_v^-}{2}\right) \right]
\nonumber \\
II+IV &:&
\int_0^{\infty} d k_v^\prime  \int_{-\infty}^0 \!\! d k_v^{\prime\prime}  
e^{-\frac{1}{2}{k_v^+}^2 D^-}
 e^{-\frac{1}{4}{k_v^-}^2 D^+} e^{ -a k_v^-}  \\
 && \times
\left[ W\left(k_v-k_v^+-\frac{k_v^-}{2}\right)
W\left(k_v^+ -k_v -\frac{k_v^-}{2}\right) +
W\left(k_v^+ + k_v +\frac{k_v^-}{2}\right)
W\left(-k_v - k_v^+ +\frac{k_v^-}{2}\right) \right]
\nonumber
\end{eqnarray}
Changing variables of integration to $k_v^+$ and $k_v^-$
\begin{eqnarray}
I+III &:& 
\int_{0}^\infty d k_v^+  e^{-\frac{1}{2}{k_v^+}^2 D^-}  e^{-2 a k_v^+} 
\int_{-2 k_v^+}^{2 k_v^+} \!\! d k_v^-
 e^{-\frac{1}{4}{k_v^-}^2 D^+}  \\
 && \times
\left[ W\left(k_v-k_v^+-\frac{k_v^-}{2}\right)
W\left(k_v^+ -k_v -\frac{k_v^-}{2}\right) +
W\left(k_v + k_v^+ +\frac{k_v^-}{2}\right)
W\left(-k_v - k_v^+ +\frac{k_v^-}{2}\right) \right]
\nonumber \\
II+IV &:&
\int_{-\infty}^\infty d k_v^+ e^{-\frac{1}{2}{k_v^+}^2 D^-}
 \int_{|2 k_v^+|}^\infty \!\! d k_v^-  
 e^{-\frac{1}{4}{k_v^-}^2 D^+} e^{ -a k_v^-}  \\
 && \times
\left[ W\left(k_v-k_v^+-\frac{k_v^-}{2}\right)
W\left(k_v^+ -k_v -\frac{k_v^-}{2}\right) +
W\left(k_v + k_v^+ +\frac{k_v^-}{2}\right)
W\left(-k_v - k_v^+ +\frac{k_v^-}{2}\right) \right]
\nonumber
\end{eqnarray}
At the end, the integrals can be combined into the main contribution and the correction that
manifests itself only when Lorentz broadening is significant.
\begin{eqnarray}
main &:&  \int_{0}^\infty d k_v^+  e^{-\frac{1}{2}{k_v^+}^2 D^-}  e^{-2 a k_v^+} 
\int_{0}^{\infty} \!\! d k_v^-e^{-\frac{1}{4}{k_v^-}^2 D^+}  \tilde W(k_v,k_v^+, k_v^-) \\
corr &&
\int_{0}^\infty d k_v^+ e^{-\frac{1}{2}{k_v^+}^2 D^-}
 \int_{2 k_v^+}^\infty \!\! d k_v^-  
 e^{-\frac{1}{4}{k_v^-}^2 D^+} \left( e^{ -a k_v^-} - e^{-2 a k_v^+} \right)  \tilde W(k_v,k_v^+, k_v^-) 
\end{eqnarray}
where symmetrized window is
\begin{eqnarray}
&&\tilde W(k_v,k_v^+, k_v^-) \equiv \\
&&W\left(k_v-k_v^+-\frac{k_v^-}{2}\right)
W\left(k_v^+ -k_v -\frac{k_v^-}{2}\right) +
W\left(k_v + k_v^+ +\frac{k_v^-}{2}\right)
W\left(-k_v - k_v^+ +\frac{k_v^-}{2}\right) + \nonumber \\
&& W\left(k_v-k_v^+-\frac{k_v^-}{2}\right)
W\left(k_v^+ -k_v -\frac{k_v^-}{2}\right) +
W\left(k_v + k_v^+ +\frac{k_v^-}{2}\right)
W\left(-k_v - k_v^+ +\frac{k_v^-}{2}\right) \nonumber
\end{eqnarray}
The final expression for $P_\tau(k_v)$ is then
\begin{eqnarray}
P_\tau(k_v)&=& \alpha^2 S \int_0^S dz (S-|z|) \xi(z) 
\int_0^\infty d k_v^+ e^{-2k_v^+a}
e^{-\frac{1}{2}{k_v^+}^2 D^-}
\int_0^{\infty} d k_v^- 
e^{-\frac{1}{4}{k_v^-}^2 D^+}
\tilde W \\
&+&
\alpha^2 S \int_0^S dz (S-|z|) \xi(z) 
\int_0^\infty \!\!\!\! d k_v^+ 
e^{-\frac{1}{2}{k_v^+}^2 D^-}
\int_{2 k_v^+}^{\infty} \!\!\!\! d k_v^- 
e^{-\frac{1}{4}{k_v^-}^2 D^+} \left(e^{-k_v^-a}-e^{-2k_v^+a} \right)
\tilde W \nonumber
\end{eqnarray}



\begin{thebibliography}{}
\bibitem[]{2069} Armstrong, J.~W., Rickett, B.~J., \& Spangler, S.~R. 1995,
                 ApJ, 443, 209
 \bibitem[]{} Ballesteros-Paredes, J.,  Klessen, R., Mac Low, M. \& Vasquez-Semadeni, E. 2006, 
 in ``Protostars and Planets V'', Reipurth, D. Jewitt, and K. Keil (eds.), 
University of Arizona Press, Tucson, 951 pp., 2007., p.63-80

\bibitem[]{207} Beresnyak, A. \& Lazarian, A., ApJL, 640, L175
\bibitem[]{2071} Beresnyak, A., Lazarian, A., Cho, J. 2005, ApJ, 624, L93
\bibitem[]{209} Biskamp, D. 2003, Magnetohydrodynamical Turbulence (Cambridge,
CUP)
 
\bibitem[]{} Brunt, C.M., Heyer, M.H., Vazquez-Semadeni, E., \& Pichardo, B. 2003,  ApJ, 595, 824-84
\bibitem[]{} Cambresy, L. 1999, {\it A}\&{\it A}, 345, 965



\bibitem[]{2072} Chepurnov, A. \& Lazarian, A. 2006a, astro-ph/0611463

\bibitem[]{} Chepurnov, A. \& Lazarian, A. 2006b, astro-ph/0611465
 

\bibitem[]{} Chepurnov, A., Lazarian, A., Stanimirovic, S.,
Peek, J., \& Heiles, C. 2006,  astro-ph/0611462

\bibitem[]{2073} Cho, J., \& Lazarian, A. 2003, \mnras, 345, 325
\bibitem[]{2074} Cho, J., \& Lazarian, A. 2004, \apj, 615, L41
\bibitem[]{2075} Cho, J., \& Lazarian, A. 2005, Theoretical and Computational 
Fluid Dynamics, 19, 127 
\bibitem[]{2077} Cho, J., Lazarian, A., Honein, A., Knaepen, B., Kassinos, S., 
\& Moin, P. 2003, ApJ, 589, L77

\bibitem[]{}  Deshpande, A.A. 2000,  MNRAS, 317, 199

\bibitem[]{} Deshpande, A.A., Dwarakanath, K.S.,  Goss, W.M., 2000,  ApJ, 543, 227


\bibitem[]{2079} Dickman, R. L., \& Kleiner, S. C. 1985, \apj, 295, 479
\bibitem[]{2080} Elmegreen, B. 2002, ApJ, 577, 206
\bibitem[]{2081} Elmegreen, B. \& Falgarone, E. 1996, ApJ, 471, 816

\bibitem[]{} Elmegreen, B. \& Scalo, J. 2004,  {\it ARA}\&{\it A}, {\bf 42},
211
 
\bibitem[]{2082} Esquivel, A. \& Lazarian, A. 2005, ApJ, 631, 320
\bibitem[]{} Esquivel, A., Lazarian, A., Horibe, S., Ossenkopf, V., Stutzki, J., Cho, J.  2007,   MNRAS,  381, 1733

\bibitem[]{2083} Esquivel, A., Lazarian, A., Pogosyan, D., \& Cho, J. 2003, 
MNRAS, 342, 325
\bibitem[]{2085} Falgarone, E. 1999,
    in {\it Interstellar Turbulence}, ed. by J. Franco,
                A. Carraminana, CUP, (henceforth
{\it Interstellar Turbulence}) p.132

\bibitem[]{} Gill, A. \& Henriksen, R.N. 1990, ApJ, 365, L27


\bibitem[]{2089} Heiles, C. \& Troland, T.H. 2004, ApJS, 151, 271
\bibitem[]{2090} Heyer, M. \& Brunt, C. 2004, ApJ, 615, 45
\bibitem[]{2091} Inogamov, N.A. \& Sunyaev, R.A. 2003, Astronomy Letters, 29, 791

\bibitem[]{} Khalil, A, Joncas, G., Nekka, F., Kestener, P., \& Arneodo, A. 
2006,  ApJS,  165, 512-550 


\bibitem[]{2092} Kim, J., \& Ryu, D. 2005, ApJ, 630, L45
\bibitem[]{2093} Kleiner, S. C., \& Dickman, R. L. 1985, \apj, 295, 466
\bibitem[]{2094} Lazarian, A. 1995, A\& A, 293, 507

\bibitem[]{} Lazarian, A. 1999, 
in {\it Plasma Turbulence and Energetic Particles in Astrophysics}, Eds. Michal Ostrowski and 
Reinhard Schlickeiser, Cracow, 28-47

\bibitem[]{2095} Lazarian, A. 2004, Journal of Korean Astronomical Society, 37, 563
\bibitem[]{2096} Lazarian, A. 2005, BAAS, 207, 119.02 
\bibitem[]{2097} Lazarian, A. 2006, Astron. Nachricht., 609, issue 5/6, 327 

\bibitem[]{2098} Lazarian, A. \& Esquivel, E. 2003, ApJ, 592, L37
\bibitem[]{2099} Lazarian, A., \& Pogosyan, D. 2000, ApJ, 537, 72
\bibitem[]{2100} Lazarian, A., \& Pogosyan, D. 2004, ApJ, 616, 943
 \bibitem[]{} Lazarian, A. \& Pogosyan, D. 2006,  ApJ, 652,
1348-1366

\bibitem[]{2101} Lazarian, A., Pogosyan, D., \& Esquivel, A.  2002, in ASP Conf. Ser. 276, Seeing Through the  Dust, ed. R. Taylor, T. L. Landecker, \& A. G. Willis (San Francisco: ASP),182
\bibitem[]{2102} Lazarian, A., Pogosyan, D., V\'{a}zquez-Semadeni, E., \& Pichardo,
B. 2001, \apj, 555, 130
\bibitem[]{} Lazarian, A., Vishanic, E., Cho, J. 2004, \apj, 603, 180
\bibitem[]{2104} Lazarian, A. \& Yan, H. 2004, in ``Astrophysical Dust''
eds. A. Witt \& B. Draine, APS, V. 309, p.479
\bibitem[]{210} Maron, J. \& Goldreich, P. 2001, ApJ, 554, 1175

\bibitem[]{} Mac Low, M.-M. \& Klessen, R.S. 2004, {\it Reviews of Modern  Physics}, { 76}, 125

\bibitem[]{2106} McKee, Christopher F.; Tan, Jonathan C. 2002, Nature, 416, 59

 \bibitem[]{} McKee, C. \& Ostriker, E. 2007, Theory of Star Formation, {\it ARA}\&{\it A}, 45, 565-687

\bibitem[]{2107} Monin, A.S. \& Yaglom, A. M. 1975, Statistical Fluid Mechanics:
Mechanics of Turbulence, Vol. 2 (Cambridge: MIT Press)
\bibitem[]{} Munch, G. 1999, in "Interstellar Turbulence", eds. J. Franco and A. Carraminana, CUP, p. 1
\bibitem[]{2109} Munch, G. 1958, Rev. Mod. Phys., 30, 1035
\bibitem[]{2110} Narayan, R., \& Goodman, J. 1989, MNRAS, 238, 963

\bibitem[]{} Ossenkopf, V., Esquivel, A., Lazarian, A. \& Stutzki, J. 2005,
{\it A}\&{\it A}, {\bf 452}, 223-236
\bibitem[]{} O'Dell C.R. \& Casteneda, H., 1987, ApJ, 317, 686


\bibitem[]{} Rosolowsky, E.W. Goodman, A. A.; Wilner, D. J.; Williams, J. P. 1999, ApJ, 524, 887

\bibitem[]{2111} Padoan, P., Goodman, A.A., \& Javela, M. 2003, ApJ, 588, 881
\bibitem[]{2112} Padoan, P., Jimenez, R., Juvela, M. \& Norlund, A. 2004, ApJ, L49

\bibitem[]{} Padoan, P., Rosolowsky, E. W., \& Goodman, A. A. 2001,
ApJ, 547, 862
\bibitem[]{} Padoan, P., Juvela, M., Kritsuk, A. \& Norman, M. 2006,  ApJ,
653, L125

\bibitem[]{2113} Pudritz, R. E. 2001, From Darkness to Light: Origin and
Evolution of Young Stellar Clusters, ASP , Vol. 243. Eds  T. Montmerle
and P. Andre. San Francisco, p.3
\bibitem[]{2116} Spangler, S.R., \& Gwinn, C.R. 1990, ApJ, 353, L29

\bibitem[]{} Spitzer, L. 1978, Physical Processes in the Interstellar Medium, Wiley-Interscience, New York


\bibitem[]{2117} Stanimirovi\'{c}, S., \& Lazarian, A., 2001, \apjl, 551, 53
\bibitem[]{2118} Stutzki, J. 2001, Astrophysics and Space Science Supplement, 277, 39
\bibitem[]{2119} Sunyaev, R.A., Norman, M.L., \& Bryan, G.L. 2003, Astronomy Letters,
29, 783
\bibitem[]{2121} von Hoerner, S. 1951, Zeitschrift f\"{u}r Astrophysics, 30, 17
\bibitem[]{2122} Wilson, O.C., Munch, G., Flather, E.M., \& Coffeen, M.F.
1959, ApJS, 4, 199

\bibitem[]{} Vazquez-Semadeni, E., Ostriker, E.C., Passot, T., Gammie, C.F.,
Stone, J.M. 2000, in {\it Protostars and Planets IV}, Tucson: University of
Arisona Press, p. 3

\end{thebibliography}
\end{document}